\documentclass{wqcd03}                 

\usepackage{txfonts}                   
\newcommand{\be}{\begin{equation}}
\newcommand{\ee}{\end{equation}}

\confname{QCD@Work 2003 - International Workshop on QCD, Conversano, Italy, 
14--18 June 2003}

\title{Study of $\theta$ dependence in Yang-Mills theories on the lattice} 

\author{Massimo~D'Elia\addressmark{a}}

\address[a]{Dipartimento di Fisica dell'Universit\`a di Genova
and INFN, Via Dodecaneso 33, I-16146 Genova, Italy}

\begin{document}

\begin{abstract}
We discuss the use of field theoretical techniques in the lattice 
determination of the free energy dependence on the $\theta$ angle 
in SU(N) Yang-Mills theories.
\end{abstract}

\maketitle


\section{Introduction}

The dependence of the free energy density $F(\theta)$ 
of Yang-Mills theories on the
$\theta$ angle is the subject of ongoing theretical debate.
$F(\theta)$ is defined as:

\noindent
\begin{center}
\be
\exp[ - V F(\theta) ] \equiv
\int [dA] \; e^{ - \int d^4 x {\cal L}(x)} \;  
 e^{ i \theta\, Q} \; 
\label{fthetadef}
\ee
\end{center}
where 
${\cal L}(x) = {1\over 4} F_{\mu\nu}^a(x)F_{\mu\nu}^a(x)$
is the usual Yang Mills lagrangian and 
$Q = \int \hbox{d}^4x \; q(x)$ is the topological charge,
with the topological
charge density $q(x)$ defined as

\noindent
\begin{center}
\be
q(x) = {g^2\over 64\pi^2} \epsilon_{\mu\nu\rho\sigma}
F_{\mu\nu}^a(x)F_{\rho\sigma}^a(x) = \partial_\mu K_\mu(x)\; ,
\label{defqx}
\ee
\end{center}
where $K_\mu(x)$ is the Chern current 

\noindent
\begin{center}
\be
 K_\mu = {g^2 \over {16 \pi^2}} \epsilon_{\mu\nu\rho\sigma}
 A^a_\nu \left(\partial_\rho A^a_\sigma - {1 \over 3}
 g f^{abc} A^b_\rho A^c_\sigma \right)
\label{eq:k}
\ee
\end{center}

The determination of $F(\theta)$ is a typical non-perturbative problem
and lattice QCD is in principle a natural tool to deal with it, but
the complex nature of the euclidean action for $\theta \neq 0$ forbids
the use of standard Monte Carlo simulations.
However many interesting physical aspects can be analyzed by studying 
$F(\theta)$ for small values of $\theta$. The terms in the Taylor expansion
of $F(\theta)$ around $\theta = 0$

\noindent
\begin{center}
\be
F(\theta) = F(0) + \sum_{k = 1}^\infty \frac{1}{k!} F^{(k)}(0) \theta^k \; 
\ee
\end{center}
are related to the connected moments of the topological 
charge distribution at $\theta = 0$:

\noindent
\begin{center}
\be
F^{(k)}(0) \equiv \frac{d^k}{d \theta^k} F(\theta) |_{\theta=0} = 
 - i^k \frac{\langle Q^k \rangle_c}{V} \; ,
\ee
\end{center}
which can be determined by standard lattice Monte Carlo simulations\footnote{
Other approaches for a determination of $F(\theta)$ have been tried,
based on a Fourier transform of the topological charge 
distribution or on extrapolations from simulations at 
imaginary values of $\theta$~\cite{ftheta-other}
}.

The quadratic term is proportional to the topological susceptibility,
 $\chi = \langle Q^2 \rangle/V$, which is expected to be $\neq 0$ to
the leading order in {$1/N_c$} in order to solve the so-called $U(1)$
problem~\cite{wit-79,ven-79}. 
It has been already  extensively studied by lattice simulations
(see Refs.~\cite{tep-00,gar-01} for recent reviews):
since the study of topological quantities on the lattice is always
non-trivial, several methods have been used: cooling, the field
theoretical method, and fermionic methods based on the index theorem,
all giving consistent results for $\chi$ and in agreement with the
Witten-Veneziano formula relating $\chi$ to the $\eta'$ mass.

It has been argued that,
for $\theta < \pi$, $F(\theta)$ is almost quadratic in
$\theta$, with $O(\theta^4)$ corrections  suppressed by powers
of $1/N_c$~\cite{wit-80,wit-98}. In order to verify this conjecture
it is interesting to obtain a lattice determination of the quartic term 
in the expansion
of $F(\theta)$ and measure its relative weight with respect
to the quadratic one. This has been done using the cooling
method~\cite{ehl-02} and the field theoretical methd~\cite{mde-03}.
In the following I will illustrate the details of the lattice
determination in the framework of the field theoretical method
and compare the results with those obtained by the cooling method.
Further details can be found in Ref.~\cite{mde-03}.

\section{The method}

On the lattice it is possible to define a discretized gauge invariant 
topological charge density operator $q_L(x)$, and a related
topological charge $Q_L = \sum_x q_L(x)$ (with the sum extended over
all lattice points), with the only requirement that, 
in the formal (na\"{\i}ve) continuum limit, 

\noindent
\begin{center}
\be
 q_L(x) {\buildrel {a \rightarrow 0} \over \sim} a^4 q(x) + O(a^6) \; ,
\label{eq:naive}
\ee
\end{center}
where $a$ is the lattice spacing.
A possible definition is

\noindent
\begin{center}
\be
q_L(x) = {{-1} \over {2^9 \pi^2}} 
\sum_{\mu\nu\rho\sigma = \pm 1}^{\pm 4} 
{\tilde{\epsilon}}_{\mu\nu\rho\sigma} \hbox{Tr} \left( 
\Pi_{\mu\nu}(x) \Pi_{\rho\sigma}(x) \right) \; ,
\label{eq:qlattice}
\ee
\end{center}
where $\Pi_{\mu\nu}(x)$ is the usual plaquette operator in the 
$\mu\nu$ plane, ${\tilde{\epsilon}}_{\mu\nu\rho\sigma}$ is the
standard Levi--Civita tensor for positive directions  and is otherwise
defined by the rule ${\tilde{\epsilon}}_{\mu\nu\rho\sigma} =
- {\tilde{\epsilon}}_{(-\mu)\nu\rho\sigma}$.

A proper renormalization must be performed when going towards 
the continuum limit, like for any other regularized operator. 
For instance, in spite of the  formal limit
in Eq.~(\ref{eq:naive}), the discretized topological charge
density renormalizes  multiplicatively~\cite{cdp-88}:

\noindent
\begin{center}
\be
 q_L(x) = Z(\beta) a^4(\beta) q(x) + O(a^6) \; , 
\ee
\end{center}
with a multiplicative renormalization constant $Z(\beta)$ which is a finite 
function of the bare coupling $\beta = 2 N / g_0^2$, approaching
1 as $\beta \to \infty$. 

Further renormalization constants, relating lattice to continuum
quantities, may appear when defining correlation functions of 
the topological charge. 
For example, in the case of the topological susceptibility, 

\noindent
\begin{center}
\be
\chi \equiv \frac{\langle Q^2 \rangle}{V} = 
\int \hbox{d}^4x \; \langle q(x) q(0)  \rangle \; ,
\label{eq:chi}
\ee
\end{center}
one in general is not guaranteed that the lattice definition

\noindent
\begin{center}
\be
\chi_L = \sum_x \langle  q_L(x) q_L(0) \rangle 
\label{lat_chi}
\ee
\end{center}
satisfy the correct continuum prescription for the contact term
arising in Eq.~(\ref{eq:chi}) as $x \to 0$, and this leads to the 
appearance of an additive renormalization, so that 
the lattice quantity $\chi_L$ is related to the continuum quantity
$\chi$ by

\noindent
\begin{center}
\be
\chi_L = Z(\beta)^2 a^4(\beta) \chi + M(\beta) \;.
\label{renchi}
\ee
\end{center}
The idea of the field theoretical method for the determination of 
$\chi$ is to compute and then subtract the renormalization constants
from the lattice quantity $\chi_L$, following Eq.~(\ref{renchi}).
The determination can be performed numerically using the so--called 
{\it heating method}~\cite{cdp-88,teper-89,cdpv-90,dv-92,acdgv-93,fp-94,add-97,add-97-2,addk-98}: 
the renormalization constants  are related to the {UV} 
fluctuations living on the scale of the lattice spacing that,
close enough to the continuum limit, 
are effectively decoupled from the topological
modes living on physical scales.
It is thus possible to create 
samples of configurations of fixed topological background with the UV
fluctuations thermalized, by { heating} a semiclassical configuration
of initial charge {Q}. Expectation values on this samples
give the necessary information:

\noindent
\begin{center}
\begin{eqnarray}
\langle Q_L \rangle &=& Z(\beta)\ Q  \nonumber \\
\langle Q_L^2 \rangle &=& Z(\beta) \; Q^2 + V \; M(\beta) \;
\label{eq13}
\end{eqnarray}
\end{center}
(this is also, in some sense, the idea at the basis of 
cooling, in which the UV fluctuations are suppressed by a process of 
local minimization of the action, without hopefully altering the background 
topological  content, in order to remove the renormalizations, 
i.e. $Z \to 1$ and $M \to 0$).

Improved {(smeared)} operators~\cite{cdpv-96} can be
used in order to reduce the renormalization effects, thus leading 
to improved estimates of $\chi$.

Let us now turn to the problem of the determination of the connected
quartic moment, 
$\langle Q^4  \rangle_c = \langle Q^4 \rangle - 3 \langle Q^2 \rangle^2 $.
It is clearly necessary to first understand how the lattice expectation
value  $\langle Q_L^4 \rangle $  renormalizes 
with respect to the continuum $\langle Q^4 \rangle $.
Apart from an obvious {$Z(\beta)$} multiplicative
renormalization, there will be additive renormalizations coming from contact
terms appearing in 

\noindent
\begin{center}
\be
\langle Q_L^4 \rangle = 
\int \hbox{d}^4x_1 \dots \hbox{d}^4x_4
\langle q_L(x_1) \dots  q_L(x_4) 
\rangle
\ee
\end{center}
as two or more charge densities come to the same point 
($x_i \sim x_j$ for some $i,j$).

It can be shown~\cite{mde-03} that a quite natural assumption for
the renormalization rule is the following

\noindent
\begin{center}
\be
\langle Q_L^4 \rangle = Z(\beta)^4 \langle Q^4 \rangle + 
M_{4,2}(\beta) \langle Q^2 \rangle + M_{4,0}(\beta)
\label{eq14}
\ee
\end{center}
This assumption can be shown to be theoretically sensible and allows 
a straightforward
extension of the {heating method} to determine the two
new renormalization constants 
{$M_{4,2}(\beta)$} and  {$M_{4,0}(\beta)$}.

Indeed, the measurement of $\langle Q^4_L \rangle$ 
on  a sample of configurations with background topological charge {$Q$}
gives 

\noindent
\begin{center}
\be \langle Q_L^4 \rangle = Z^4(\beta) \; Q^4  + 
M_{4,2}(\beta) \; Q^2 + M_{4,0}(\beta)
\label{eq15}
\ee
\end{center}
and if the measurement is repeated in at least two different 
topological sectors,
sufficient constraints are obtained to determine 
{$M_{4,2}(\beta)$} and  {$M_{4,0}(\beta)$}. If more
topological sectors are used, there is an excess of constraints 
which can be exploited as a non-trivial test of the method.

In the general case of the 
$n$-th order correlation function, a natural extension of 
Eq.~(\ref{eq14}) is the following:

\noindent
\begin{center}
\be
\langle Q_L^n \rangle = Z^n \langle Q^n \rangle + 
\sum_{h=1}^{n/2} 
M_{n,n-2h} \langle Q^{n-2h} \rangle \; ,
\label{ren-rule2}
\ee
\end{center}
and its validity can be discussed along the same lines as 
for $n = 4$~\cite{mde-03}.

\section{Results}

The method has been applied to the case of 
{ $SU(3)$} pure gauge theory, on a {$16^4$ lattice}
at {$\beta = 6.1$}, using {1--smeared} and {2--smeared}
operators~\cite{cdpv-96,add-97}.

We have collected five different samples of configurations,
one thermalized in the $Q = 0$ sector (around the zero field
configuration), two in the $Q = 1$ sector (thermalized around two different
semiclassical configurations of topological charge one) and two
in the $Q = 2$ sector (thermalized around two different
semiclassical configurations of topological charge two).
The semiclassical configurations have been obtained by extracting thermalized
configurations with non-trivial topology 
from the equilibrium ensemble at $\beta = 6.1$ and then minimizing
their action by a usual cooling technique.
All the five samples have been obtained by performing about 3000
heating trajectories around the semiclassical configurations, each
trajectory consisting of 90 heating steps; 
6 straight cooling steps have been applied on heated configurations
to check that their background topological content did not change.

We have then measured the expectation values
$\langle Q_L^2 \rangle$, $\langle Q_L^4 \rangle$, 
and also $\langle Q_L \rangle/Q$ where $Q \neq 0$,  over the 
five samples.
We have reported the results in Table 1 for the 1-smeared operator 
and in Table 2 for the 2-smeared operator: 
expectation values obtained on samples with the same $Q$ turned
out to be equal within errors, as they should, and we have reported
in the tables only their weighted averages. 
Those data have then been used to perform a fit to
Eqs.~(\ref{eq13}) and (\ref{eq15}) obtaining finally the values 
of the renormalization constants reported in Table 3.

\begin{center}
\begin{table}[ph]
\caption{Expectation values measured in different 
topological sectors for the 1-smeared operator.}
\label{table1}
\vspace{5pt}
{
\begin{center}
\begin{tabular}{|c|r|r|r|}
\hline
{} &{} &{} &{}\\[-1.5ex]
$Q$ & $Z = \langle Q_L \rangle/Q$ &  $\langle Q_L^2 \rangle$ & $\langle Q_L^4 \rangle$\\[1ex]
\hline
{} &{} &{} &{}\\[-1.5ex]
0 & -   & 0.311(12)  & 0.290(20)\\[1ex] 
1 & 0.416(6) & 0.4785(60) & 0.630(15)\\[1ex] 
2 & 0.413(5) & 0.9626(80) & 1.973(50)\\[1ex] 
\hline
\end{tabular}
\end{center}
}
\vspace*{-1pt}
\end{table}
\end{center}

\begin{center}
\begin{table}[ph]
\caption{Expectation values measured in different 
topological sectors for the 2-smeared operator.}
\label{table2}
\vspace{5pt}
{
\begin{center}
\begin{tabular}{|c|r|r|r|}
\hline
{} &{} &{} &{}\\[-1.5ex]
$Q$ & $Z = \langle Q_L \rangle/Q$ &  $\langle Q_L^2 \rangle$ & $\langle Q_L^4 \rangle$\\[1ex]
\hline
{} &{} &{} &{}\\[-1.5ex]
0 & - & 0.208(10)& 0.124(10)\\[1ex]
1 & 0.544(5) & 0.489(5) & 0.556(12)\\[1ex] 
2 & 0.542(4) & 1.314(8) & 2.77(6)\\[1ex]
\hline
\end{tabular}
\end{center}
}
\vspace*{-1pt}
\end{table}
\end{center}

The equilibrium values $\langle Q_L^2 \rangle$ and 
$\langle Q_L^4 \rangle$, which are reported in Table 4, have been measured
on a sample of 300K configurations separated by five updating cycles,
each composed of a mixture of 4 over-relaxation + 1 heat-bath updating sweeps;
the reported errors have been estimated by a standard blocking technique.

Using Eqs.~(\ref{renchi}) and (\ref{eq14}) we can  compute 
$\langle Q^2 \rangle$ and $\langle Q^4 \rangle$, obtaining the results
reported in Table 4.  It is interesting to notice that the
values obtained for the 1-smeared operator and for the 2-smeared operator
are in good agreement, as they should, confirming the 
correctness of the method.

We can finally determine $\langle Q^4 \rangle_c = 
\langle Q^4 \rangle - 3 \langle Q^2 \rangle^2$, obtaining
$\langle Q^4 \rangle_c = 0.32 \pm 1.80$ for the 1-smeared and
$\langle Q^4 \rangle_c = 0.66 \; \pm \; 0.90$ for the 2-smeared operator,
leading to $b_2 = -0.012(62)$ and $b_2 = -0.024(32)$ for the 
1-smeared and 2-smeared operator respectively, 
in agreement with the determination reported in Ref.~\cite{ehl-02}.

\begin{center}
\begin{table}[ph]
\caption{Values of the renormalization constants
obtained respectively for the 1-smeared and 2-smeared operators,
by using the results reported in tables 1 and 2 and
performing a best fit to Eqs.~(\ref{eq13}) and (\ref{eq15}).}
\label{table3}
\vspace{5pt}
{
\begin{center}
\begin{tabular}{|r|r|r|r|}
\hline
{} &{} &{} &{}\\[-1.5ex]
 $Z$ &  $V M$ & $M_{4,0}$  &  $M_{4,2}$\\[1ex]
\hline
{} &{} &{} &{}\\[-1.5ex]
  0.414(4) & 0.315(6) & 0.336(16) & 0.289(16)\\[1ex] 
  0.543(5) & 0.211(5) & 0.377(15) & 0.124(9)\\[1ex]
\hline
\end{tabular}
\end{center}
}
\vspace*{-1pt}
\end{table}
\end{center}

\begin{center}
\begin{table}[ph]
\caption{Expectation values measured at equilibrium
and results obtained for the renormalized quantities,
respectively for the 1-smeared and 2-smeared operators.}
\label{table4}
\vspace{5pt}
{
\begin{center}
\begin{tabular}{|r|r|r|r|}
\hline
{} &{} &{} &{}\\[-1.5ex]
$\langle Q_L^2 \rangle$ & $\langle Q_L^4 \rangle$  & $\langle Q^2 \rangle$ & $\langle Q^4 \rangle$\\[1ex]
\hline
{} &{} &{} &{}\\[-1.5ex]
 0.7121(38)  & 1.548(18) & 2.312(72) & 16.4 $\pm$ 1.8\\[1ex] 
0.8776(60) & 2.368(36) & 2.262(41) & 16.02(72)\\[1ex]
\hline
\end{tabular}
\end{center}
}
\vspace*{-1pt}
\end{table}
\end{center}

\section{More on the renormalization effects}

The renormalization constants $Z$, $M_{n,m}$ ($m < n$) which, for a 
given lattice discretization $Q_L$, appear in Eq.~(\ref{ren-rule2}),
are in principle independent of each other, or at least no simple
relation exists among them, unless some further hypothesis 
can be done about the nature of the UV fluctuations which are responsible for
the renormalizations. We will propose and test
an ansatz which will greatly simplify the structure of the renormalization
constants and will lead to a renormalization formula which directly 
involves the connected  correlation functions, thus allowing a more 
precise determination of $b_2$.

An hypothesis about the nature of the UV fluctuations has been done in 
Refs.~\cite{teper-89,dv-92}, where it was assumed that the 
discretized topological charge density can be expressed as

\noindent
\begin{center}
\be
q_L(x) \simeq [Z + \zeta (x)] q(x) + \eta (x) \; ,
\label{hypo1}
\ee
\end{center}
where $q(x)$ is a background topological charge density
which is determined by physical fluctuations on the scale
of the correlation length $\xi$, whereas $\zeta(x)$ and $\eta(x)$ 
are random variables with zero averages which are determined
by the short range UV fluctuations and, at least in the continuum
limit, are expected to be decoupled from $q(x)$, i.e.
$\langle \zeta(x) q(x) \rangle = \langle \eta(x) q(x) \rangle = 0$.
Summing Eq.~(\ref{hypo1}) over all lattice points, the following
relation follows for the lattice topological charge $Q_L$:

\noindent
\begin{center}
\be
Q_L = Z \; Q + \sum_x \zeta(x) q(x) + \eta \; ,
\label{hypo2}
\ee
\end{center}
where $\eta =  \sum_x \eta(x)$. We now make the further assumption that
the term $\sum_x \zeta(x) q(x)$ in Eq.~(\ref{hypo2}) can be neglected, 
configuration by configuration. This is not  
unreasonable, in view of the fact that $q(x)$ and $\zeta (x)$ are
decoupled from each other.
We will thus assume that 

\noindent
\begin{center}
\be
Q_L =  Z \; Q + \eta \; , 
\label{ansatz}
\ee
\end{center}
where $\eta$ is a random noise with zero average which is 
stochastically independent of $Q$. 

This assumption has relevant consequences for the structure of the 
renormalization constants. Indeed, using the hypothesis that 
$Q$ and $\eta$ are stochastically independent variables and that they are
both evenly distributed around zero, it is easy to verify that the 
general renormalization formula holds:

\noindent
\begin{center}
\be
\langle Q_L^n \rangle = \sum_{h = 0}^{n/2} {n \choose 2h} Z^{n - 2h}  
\langle Q^{n - 2h} \rangle \langle \eta^{2h} \rangle \; ,
\label{ren-rule-2}
\ee
\end{center}
so that the renormalization relation for $\langle Q_L^n \rangle$
is described only in terms of $Z$ and of the correlation
functions of the noise $\eta$. In particular we have
$M_{n,m} = {n \choose m} Z^m \langle \eta^{n-m} \rangle$,
a relation that should be verified on numerical data if our ansatz in 
Eq.~(\ref{ansatz}) is correct. From the data in Table 3 it can 
be checked that indeed $M_{4,2} = 6 Z^2 \langle \eta^2 \rangle = 
6 Z^2 V M$, but we will now proceed further and check
the validity of Eq.~(\ref{ren-rule-2}) up to $n = 6$.
The correlation functions of $\eta$ can be determined by the heating
method using the analogous of Eq.~(\ref{eq15}), which
up to $n = 6$ reads:

\noindent
\begin{center}
\begin{eqnarray}
\langle Q_L^2 \rangle &=& Z^2 Q^2 + \langle \eta^2 \rangle
\nonumber \\
\langle Q_L^4 \rangle &=& Z^4 Q^4  + 
6 Z^2 Q^2  \langle \eta^2 \rangle + \langle \eta^4 \rangle
\nonumber \\
\langle Q_L^6 \rangle &=& Z^6 Q^6 + 
15 Z^4 Q^4  \langle \eta^2 \rangle + \nonumber \\ 
&& + 15 Z^2 Q^2  \langle \eta^4 \rangle + \langle \eta^6 \rangle \; .
\label{heat-system-2}
\end{eqnarray}
\end{center}
Using the values for $\langle Q_L^2 \rangle$, $\langle Q_L^4 \rangle$ and
$\langle Q_L^6 \rangle$ obtained in the sectors with $Q = 0,1,2$ 
we have performed a best fit to 
Eqs.~(\ref{heat-system-2}), obtaining the best fit values reported in Table 5 
with  $\chi^2/{\rm d.o.f.} \simeq 0.34$ for the 1-smeared operator and 
$\chi^2/{\rm d.o.f.} \simeq 0.23$ for 2-smeared operator.
The fact that the values for 
$\langle Q_L^2 \rangle$, $\langle Q_L^4 \rangle$ and
$\langle Q_L^6 \rangle$ obtained in the various 
sectors can be fitted by the simple relations in Eq.~(\ref{heat-system-2})
is a confirmation of the validity of the ansatz 
in Eq.~(\ref{ansatz}).

Assuming that Eq.~(\ref{ansatz}) is valid, it is 
possible to write a renormalization relation which involves directly
the connected correlation functions. Indeed, it is a general rule
that the connected correlation functions of a stochastic variable
($Q_L$ in our case), which is the sum of two variables which are 
stochastically independent of each other ($Z Q$ and $\eta$ in our case),
are the sum of the corresponding connected correlation functions, i.e.

\noindent
\begin{center}
\be
\langle Q_L^n \rangle_c = Z^n \langle Q^n \rangle_c + 
\langle \eta^n \rangle_c \; .
\label{simple-ren}
\ee
\end{center}

Therefore in order to compute $\langle Q^n \rangle_c$
we need to know, apart from $Z$, only one renormalization constant,
$\langle \eta^n \rangle_c$, which can be easily measured 
by computing $\langle Q_L^n \rangle_c$ on the sample of configurations
in the $Q = 0$ sector. 
$\langle Q^n \rangle_c$ is then given by

\noindent
\begin{center}
\be
\langle Q^n \rangle_c = 
\frac{\langle Q_L^n \rangle_c - \langle \eta^n \rangle_c}{Z^n} \; ,
\label{simple}
\ee
\end{center}
where $\langle Q_L^n \rangle_c$ is measured on the ensemble of 
configurations at equilibrium.

We have computed 
$\langle Q_L^4 \rangle_c = \langle Q_L^4 \rangle - 3 \langle Q_L^2 \rangle^2 $ 
on our equilibrium configurations at $\beta = 6.1$, obtaining  
$\langle Q_L^4 \rangle_c = 0.026(7)$ for the 1-smeared operator
and $\langle Q_L^4 \rangle_c = 0.057(13)$ for the 2-smeared operator.
We have then computed $\langle Q_L^4 \rangle_c$ on our sample of configurations
thermalized in the $Q = 0$ sector at $\beta = 6.1$ , obtaining
$\langle \eta^4 \rangle_c = \langle \eta^4 \rangle - 3 \langle \eta^2 \rangle^2 = 0.006(4)$ for the 1-smeared operator
and $\langle \eta^4 \rangle_c = 0.001(2)$ for the 2-smeared operator.
In both cases (equilibrium and $Q = 0$) errors have been estimated by standard
 jackknife techniques.

By using Eq.~(\ref{simple}) and the values for $Z$ and $\langle Q^2 \rangle$
previously obtained, we have obtained $\langle Q^4 \rangle_c = 0.68(24)$,
$b_2 = -0.024(9)$ for the 1-smeared operator and 
$\langle Q^4 \rangle_c = 0.66(15)$, $b_2 = -0.024(6)$ 
for the 2-smeared operator.

By making use of the ansatz in Eq.~(\ref{ansatz}) we have thus
made determinations which are much more precise than those obtained
in Section 3. The reason is that Eq.~(\ref{simple-ren}) allows
to relate $\langle Q^n \rangle_c$ directly to the connected correlation
functions of the discretized lattice topological charge, with only two
renormalization constants involved: this greatly simplifies
computations and error propagation, thus leading to improved
estimates. 
We notice that most of the error in the final
determination of $\langle Q^4 \rangle_c$ and  $b_2$ comes
from the determination of $\langle Q_L^4 \rangle_c$ at equilibrium,
which is also the most expensive part of the computation in terms
of CPU time. The renormalization procedure is thus completely under
control and numerically non expensive.

We have also made a determination of $b_2$ at $\beta = 6.0$,
again on a $16^4$ lattice.
On a sample of about 300K configurations obtained at equilibrium and
using the same algorithm as for $\beta = 6.1$ we have obtained,
for the 2-smeared operator, $\langle Q_L^2 \rangle = 1.377(7)$,  
$\langle Q_L^4 \rangle_c = 0.052(23)$. On a sample of configurations
thermalized in the $Q = 0$ topological sector by performing about
3000 heating trajectories, each composed of 90 heating steps, we
have obtained, for the 2-smeared operator,  
$\langle \eta^2 \rangle = 0.308(10)$ and 
$\langle \eta^4 \rangle_c = 0.002(3)$. From these data, using 
the value $Z(\beta = 6.0) = 0.51(2)$ reported for the 2-smeared operator
in Ref.~\cite{add-97}, we obtain $b_2 = -0.015(8)$, which is consistent
with the value obtained at $\beta = 6.1$.

Let us close noticing that the value obtained for $\langle \eta^4 \rangle_c$
is very small and compatible with zero for both the 1-smeared and 
the 2-smeared operator. We have also measured $\langle \eta^6 \rangle_c$
 on the sample of configurations at $Q = 0$ obtaining
$\langle \eta^6 \rangle_c = 0.001(8)$ for the 1-smeared and
$\langle \eta^6 \rangle_c = 0.0005(14)$ for 2-smeared operator ($\beta = 6.1$),
so that $\eta$ behaves with a good approximation as a pure gaussian noise.

\begin{center}
\begin{table}[ph]
\caption{Values of the renormalization constants,
respectively for the 1-smeared and 2-smeared operators,
obtained by performing a best fit to Eq.~(\ref{heat-system-2}).}
\label{table5}
\vspace{5pt}
{
\begin{center}
\begin{tabular}{|r|r|r|r|}
\hline
{} &{} &{} &{}\\[-1.5ex]
 $Z$ & $\langle \eta^2 \rangle$ & $\langle \eta^4 \rangle$ & $\langle \eta^6 \rangle$\\[1ex]
\hline
{} &{} &{} &{}\\[-1.5ex]
 0.415(4) & 0.315(6) & 0.298(11) & 0.462(37)\\[1ex]
 0.542(4) & 0.211(5) & 0.129(6) & 0.131(17)\\[1ex]
\hline
\end{tabular}
\end{center}
}
\vspace*{-1pt}
\end{table}
\end{center}

\section{Conclusions}

We have discussed the extension of
the field theoretical method, already used for
the lattice determination of the topological 
susceptibility, to the computation of 
further terms in the expansion of the 
ground state energy $F(\theta)$ around $\theta = 0$.

We have presented numerical results regarding SU(3)
pure gauge theory, providing a determination
of the fourth order term in the expansion of 
$F(\theta)$ around $\theta = 0$.
Our determination is in agreement with that
obtained in Ref.~\cite{ehl-02} via the cooling method,
and  confirms that fourth order corrections to  
the simple $\theta^2$ behaviour of $F(\theta)$ around $\theta = 0$
are small already for $N_c = 3$.

\section*{Acknowledgments}

I am indebted to B.~All\'es, L.~Del~Debbio, A.~Di~Giacomo 
and E.~Vicari for useful comments and discussions and
to the computer center of ENEA for providing time on
their QUADRICS machines. I would like to thank the
workshop organizers for their hospitality and for the
warm atmosphere provided at the workshop. 
This work has been partially supported by MIUR.


\begin{thebibliography}{9}


\bibitem{ftheta-other} S.~Olejnik and G.~Schierholz,
Nucl.\ Phys.\ Proc.\ Suppl.\  {\bf 34} (1994) 709; 
J.~C.~Plefka and S.~Samuel,
Phys.\ Rev.\ D {\bf 56} (1997) 44; 
R.~Burkhalter, M.~Imachi, Y.~Shinno and H.~Yoneyama,
Prog.\ Theor.\ Phys.\  {\bf 106} (2001) 613; 
V.~Azcoiti, G.~Di Carlo, A.~Galante and V.~Laliena,
Phys.\ Rev.\ Lett.\  {\bf 89} (2002) 141601; 
V.~Azcoiti, G.~Di Carlo, A.~Galante and V.~Laliena,
Phys.\ Lett.\ B {\bf 563} (2003) 117.

\bibitem{wit-79} E.~Witten, Nucl. Phys. {\bf B 156} (1979) 269.

\bibitem{ven-79} G.~Veneziano, Nucl. Phys. {\bf B 159} (1979) 213.

\bibitem{tep-00} 
M. Teper, Nucl. Phys. (Proc. Suppl.) {\bf 83}, 146 (2000). 

\bibitem{gar-01}
M.~Garc\'{\i}a P\'erez,
Nucl.\ Phys.\ (Proc.\ Suppl.)  {\bf 94}, 27 (2001).

\bibitem{wit-80} E.~Witten,  Ann. Phys. (NY) {\bf 128} (1980) 363.

\bibitem{wit-98} E.~Witten, Phys. Rev. Lett. {\bf 81} (1998) 2862.

\bibitem{ehl-02} L.~Del Debbio, H.~Panagopoulos and E.~Vicari, 
JHEP 0208:044,2002.

\bibitem{mde-03} M. D'Elia, Nucl.\ Phys.\ B {\bf 661} (2003) 139 and
arXiv:hep-lat/0302007.



\bibitem{cdp-88} M. Campostrini, A. Di Giacomo and H. Panagopoulos,
Phys. Lett. {\bf B 212} (1988) 206.

\bibitem{teper-89} M. Teper, Phys. Lett. {\bf B 232} (1989) 227.

\bibitem{cdpv-90} M. Campostrini, A. Di Giacomo H. Panagopoulos and
E. Vicari, Nucl.\ Phys.\ B {\bf 329} (1990) 683.

\bibitem{dv-92} A. Di Giacomo and E. Vicari, 
Phys. Lett. {\bf B 275} (1992) 429.

\bibitem{acdgv-93} B. All\'es, M. Campostrini, A. Di Giacomo, Y. G\"und\"u\c c
and E. Vicari, Phys. Rev. {\bf D 48} (1993) 2284.

\bibitem{fp-94} F. Farchioni and A. Papa, 
Nucl.\ Phys.\ B {\bf 431} (1994) 686.

\bibitem{add-97}
B.~All\'es, M.~D'Elia and A.~Di Giacomo,
Nucl.\ Phys.\ B {\bf 494} (1997) 281.

\bibitem{add-97-2}
B.~All\'es, M.~D'Elia and A.~Di Giacomo,
Phys. Lett. {\bf B 412} (1997) 119.

\bibitem{addk-98} B.~All\'es, M.~D'Elia, A.~Di Giacomo and R. Kirchner,
Phys. Rev. {\bf D 58}:114506, (1998).

\bibitem{cdpv-96} C. Christou, A. Di Giacomo, H. Panagopoulos and E. Vicari,
Phys. Rev. {\bf D 53} (1996) 2619.

\end{thebibliography}
\end{document}